\documentclass[12pt]{iopart}
\usepackage{graphicx}
\usepackage{iopams}  
\usepackage{hyperref}
\usepackage{xcolor, soul}
\hypersetup{
    colorlinks=true,
    citecolor=red
}  

\begin{document}

\title{Mesoscopic transport in KSTAR plasmas: avalanches and the $E \times B$ staircase}

\author{Minjun J. Choi$^1$, Jae-Min Kwon$^1$, Lei Qi$^1$, P. H. Diamond$^2$, T. S. Hahm$^3$, Hogun Jhang$^1$, Juhyung Kim$^1$, M. Leconte$^1$, Hyun-Seok Kim$^1$, Jisung Kang$^1$, Byoung-Ho Park$^1$, Jinil Chung$^1$, Jaehyun Lee$^1$, Minho Kim$^1$, Gunsu S. Yun$^4$, Y. U. Nam$^1$, Jaewook Kim$^1$, Won-Ha Ko$^1$, K. D. Lee$^1$, J. W. Juhn$^1$ and the KSTAR team}

\address{$^1$ Korea Institute of Fusion Energy, Daejeon 34133, Republic of Korea}
\address{$^2$ University of California, San Diego, La Jolla, California 92093-0424, U.S.A.}
\address{$^3$ Seoul National University, Seoul 08826, Republic of Korea}
\address{$^4$ Pohang University of Science and Technology, Pohang, Gyungbuk 37673, Republic of Korea}

\ead{mjchoi@kfe.re.kr}
	
\begin{abstract}

The self-organization is one of the most interesting phenomena in the non-equilibrium complex system, generating ordered structures of different sizes and durations. 
In tokamak plasmas, various self-organized phenomena have been reported, and two of them, coexisting in the near-marginal (interaction dominant) regime, are avalanches and the $E \times B$ staircase.
Avalanches mean the ballistic flux propagation event through successive interactions as it propagates, and the $E \times B$ staircase means a globally ordered pattern of self-organized zonal flow layers. 
Various models have been suggested to understand their characteristics and relation, but experimental researches have been mostly limited to the demonstration of their existence.
Here we report detailed analyses of their dynamics and statistics and explain their relation. 
Avalanches influence the formation and the width distribution of the $E \times B$ staircase, while the $E \times B$ staircase confines avalanches within its mesoscopic width until dissipated or penetrated.  
Our perspective to consider them the self-organization phenomena enhances our fundamental understanding of them as well as links our findings with the self-organization of mesoscopic structures in various complex systems. 

\end{abstract}

\maketitle

\section{Introduction}

Thermodynamic non-equilibrium complex systems~\cite{Nicolis1989, GoldenfeldScience1999, Sanchez2018}, composed of a great number of interacting elements, feature the self-organization phenomena, generating structures whose distributions of size and duration are rarely normal and often have a long tail. 
Tokamak plasmas can be considered one of the thermodynamic non-equilibrium complex systems.
The central region of the plasma is heated and fueled by external devices, and the edge region is attached to any form of sink.
Plasmas reach a stationary non-equilibrium state with the finite heat flux and pressure gradient between the hot center and the cold edge.
The pressure gradient serves as a drive for various plasma instabilities. 
Microscopic ($\sim$ the gyro-radius $\rho$) instabilities, or simply plasma turbulence, can interact over the long range~\cite{HahmJKPS2018}, and plasmas can exhibit emergent features of a complex system.

Plasmas as a complex system imply that a useful picture of fluctuation loss by plasma turbulence, i.e. the local diffusive paradigm, may need to be limited.
Understanding and controlling transport have been of primary importance for the success of tokamaks.  
In case that macroscopic ($\sim$ the plasma radius $a$) magnetohydrodynamic (MHD) instabilities are suppressed, fluctuation loss dominate the transport.
The local diffusive paradigm, relying on the existence of the characteristic transport scales and the locality of the flux-gradient relation, has provided a practical method to assess fluctuation loss. 
%Plasmas as a complex system imply that a useful picture of fluctuation loss, i.e. the local diffusive paradigm, may need to be limited.
%The local diffusive paradigm relies on the existence of the characteristic transport scales and the locality of the flux-gradient relation. 
%However, microscopic instabilities in tokamak plasmas can interact over the long range~\cite{HahmJKPS2018}, and plasmas can exhibit emergent features of a complex system.
However, various self-organized structures have been reported in tokamak plasmas~\cite{ItohPFR2009}, and two of them, which coexist in the near-marginal regime, are avalanches and the $E \times B$ staircase. 
Avalanches~\cite{DiamondPoP1995, HahmJKPS2018} mean the ballistic flux propagation events through radially successive (non-local~\cite{DifPRE2010}) interactions~\cite{CarrerasPoP1996, IdomuraNF2009, KuNF2009, SarazinNF2010, PolitzerPRL2000, PanNF2015, VanCompernollePoP2017, Choi:2019wy, Zhang:2019dq, KinNF2023}.
They are named after the avalanche of the self-organized criticality (SOC) system~\cite{BakPRL1987, HwaPRA1992, Jensen1998, Livi2017, Sanchez2018}, because they are fast relaxation events of various sizes, exhibiting the power-law spectrum (long time correlation) and the large Hurst exponent~\cite{SanchezPPCF2015}. 
The $E \times B$ staircase~\cite{DifPRE2010, DifNF2017} means globally self-organized zonal flow layers separated by a mesoscopic ($\rho < \Delta < a$) width~\cite{DifPRE2010, GhendrihEPJD2014, Villard:2014bs, DominskiPoP2015, WangNF2018, QiNF2019, QiNF2021, RathPoP2021, DifPRL2015, HornungNF2017, Choi:2019wy, Ashourvan:2019ek, vanMilligenE2019, LiuPoP2021, Kosuga:2013io, ItohPPCF2016, AshourvanPRE2016, GuoPPCF2019, SasakiPoP2021, GarbetPoP2021, LecontePoP2021, LecontePoP2022, YanNF2022}.
It is named after the potential vorticity staircase observed in geophysical fluid~\cite{McIntyreNature1983}, reflecting the similarity of the zonal flow generation via the potential vorticity flux (the Reynolds force) in two systems. 
% and radial force balance
The $E \times B$ staircase acts like mini permeable transport barriers, confining small avalanches within its mesoscopic width and producing a staircase-like pressure profile corrugation~\cite{Choi:2019wy, Ashourvan:2019ek, LiuPoP2021}. 
When avalanches and the $E \times B$ staircase coexist, they mostly result in a transport of a given mesoscopic width $\Delta$~\cite{DifNF2017}.
However, the staircase is permeable to the large but rare avalanche and its width distribution has a long tail (even the width can change in time considerably)~\cite{GhendrihEPJD2014, DifNF2017}, making the transport a process of mesoscopic range rather than a definite scale. 
% On the other hand, they exhibit a kind of critical behavior, meaning that they form by the rapid propagation of an initially localized seed perturbation. 
These self-organized structures by their inherent nature would question the validity of transport frameworks based on the local diffusive paradigm in the near-marginal regime, but as shown in the recent sophisticated model comparison~\cite{GillotPPCF2023} it would be also an opportunity to test and improve physics models.

Despite their importance, experimental observations are reported limitedly and somewhat confusingly. 
This may be partly due to our insufficient understanding of these structures. 
The numerical simulation found that the $E \times B$ staircase is most prominent in the near-marginal (interaction dominant) regime~\cite{DifNF2017}. 
The near-marginal regime means a regime close to the nonlinear threshold~\cite{GillotPPCF2023}, but it does not necessarily mean the least external drive. 
The input of low-entropy energy is rather the key to the emergence of such a macroscopic order in complex systems~\cite{Nicolis1989}. 
Considering these structures the self-organization in non-equilibrium complex systems, they can be regarded as efficient ways to release (avalanches) or extract/dissipate (staircase) the excess energy for the system to reach a stationary state.
Indeed, in KSTAR experiments (see Methods~\ref{sec:avalregime})~\cite{Choi:2019wy}, they are observed with some amount of external heating under the unfavorable shape (a strong sink) to avoid a typical (but MHD-vulnerable) structure formation, i.e. the edge transport barrier~\cite{WagnerPRL1982, ConnorPPCF2000}. 

Another hindrance may be the diagnostic limitation in core fusion plasmas. 
For the $E \times B$ staircase, observations have been reported in toroidal plasmas based on its footprints on some measurable quantities.  
In Tore Supra~\cite{DifPRL2015, HornungNF2017} and HL-2A~\cite{LiuPoP2021}, the radial variation of the turbulence correlation length and the corresponding turbulence eddy tiltings are provided as evidence for the shear flow layers, the $E \times B$ staircase. 
In TJ-II~\cite{vanMilligenPoP2018}, W7-X~\cite{vanMilligenNF2018} and JET~\cite{vanMilligenE2019}, the radial transfer of information in fluctuation is analyzed to identify the radially localized mini barriers at which the information is trapped relatively longer~\cite{vanMilligenE2019}. 
In KSTAR~\cite{Choi:2019wy}, DIII-D~\cite{Ashourvan:2019ek} and HL-2A~\cite{LiuPoP2021}, the corrugation of temperature/density profiles is provided, demonstrating the existence of multiple mesoscopic transport barriers. 
It is noteworthy that the co-existence of the $E \times B$ staircase and avalanches was only reported in the KSTAR experiment~\cite{Choi:2019wy}. 

We report detailed dynamics and statistics of avalanches and the $E \times B$ staircase in KSTAR plasmas to understand their characteristics and mutual relation.  
%The rest of this paper is organized as follows. 
%In Section~\ref{sec:kstaravst}, it will be explained how avalanches and the $E \times B$ staircase are identified by their footprints on the electron temperature. 
Since the ion temperature gradient (ITG) mode is dominant in the KSTAR avalanche/staircase plasmas, the ion channel would be of more interest. 
However, the current resolution of the ion diagnostics is not sufficient to study those structures of various scales, though big events are also evident in the ion channel.
Therefore, their identification and characterization are based on their footprints on the electron temperature ($T_\mathrm{e}$) and density ($n_\mathrm{e}$) (see Methods~\ref{sec:kstaravst}).
%For example, avalanches appear as bumps and voids in $T_\mathrm{e}$, propagating ballistically, and the $E \times B$ staircase manifests as the $T_\mathrm{e}$ profile corrugation.  
This may be justified by that the analysis results are not inconsistent with the expected results from theory or simulation. % in Section~\ref{sec:results}
For avalanches, their ballistic propagation behavior is shown and their propagation speed is compared with a model prediction~\cite{GurcanPoP2005}.
The formation of a radially extended eddy structure is observed in the edge region with the long range propagation of avalanches.  
Avalanches have the power-law distribution of the pseudo-size. 
For the $E \times B$ staircase, its initial formation from a seed perturbation and dynamical transformations are observed. 
Its avalanche stopping capability~\cite{GhendrihEPJD2014} is demonstrated, and a radially restricted eddy structure within the staircase width is observed. 
The three-wave coupling between the zonal component and the broadband arises near both ends of the structure. 
Distributions of the staircase characteristics seem to have a long tail, and the width distribution of avalanche-associated-staircases fits with a Fr\'echet distribution as in the simulation~\cite{DifNF2017}.  
We find that avalanches can provide a seed perturbation for the staircase formation, suggesting one way to understand its particular width distribution. 
A condition for the staircase formation is discussed by incorporating our observations and previous analyses~\cite{Kosuga:2013io, WangNF2018}. 
The staircase confines small avalanches within its mesoscopic width, working as transport barriers, but it is penetrated by large and rare avalanches. 
Besides the intellectual interest in these intricate phenomena, we note that the improved understanding of the near-marginal regime transport would be critical to advancing the predictive capability of plasma transport models~\cite{GillotPPCF2023}.
Beyond the comparison of profiles or transport coefficients, characteristics and distributions of transport players can be compared, enabling the higher level validation of models.

% \section{Analyses of avalanches \label{sec:aval}}
 \section{Results \label{sec:results}}

\subsection{The ballistic propagation of avalanches \label{sec:avalball}}

\begin{figure}[t]
\includegraphics[keepaspectratio,width=0.6\textwidth]{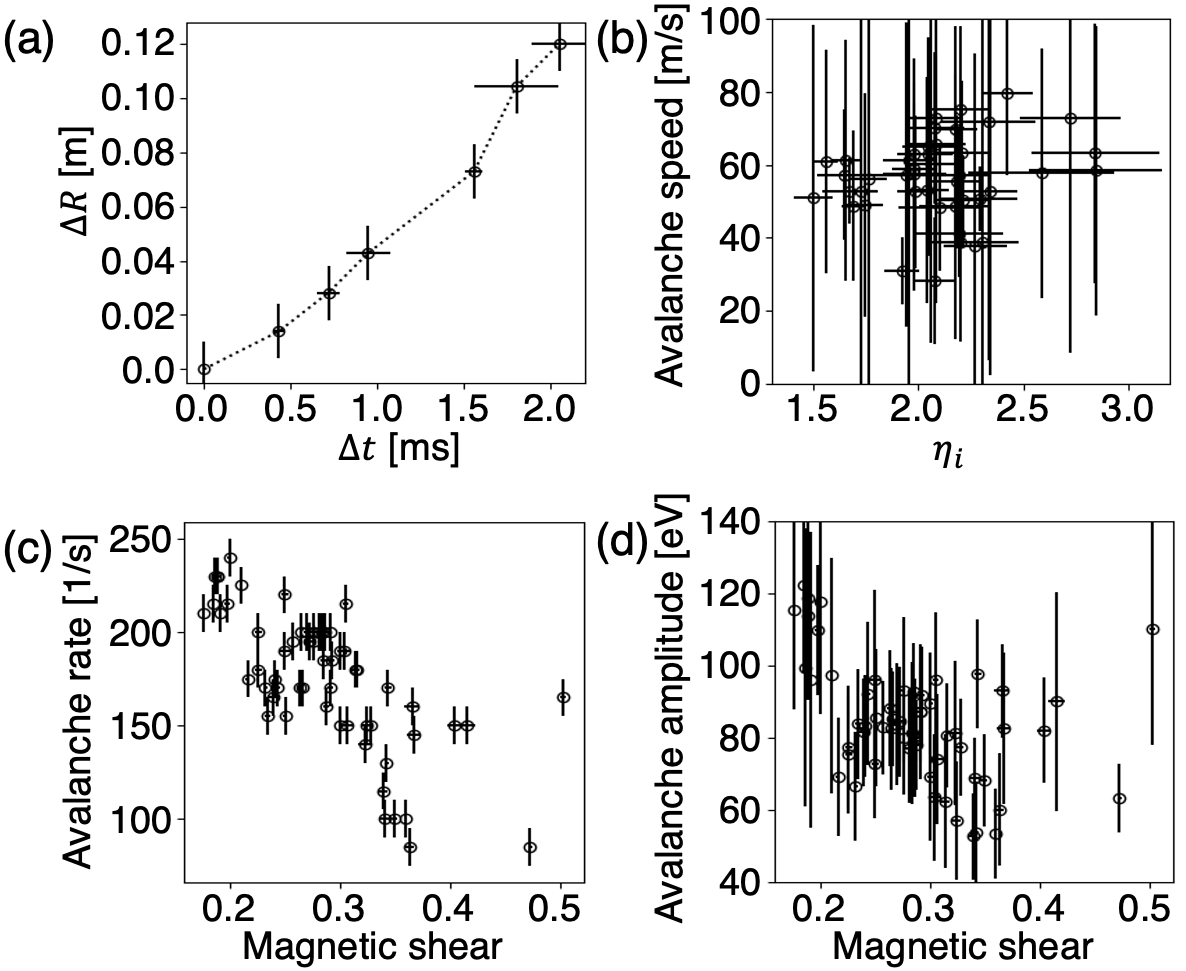}
\caption{(a) The channel distance $\Delta R$ versus the maximum correlation time lag $\Delta t$. (b) The average avalanche propagation speed versus $\eta_i \equiv L_n/L_{T_\mathrm{i}}$. (c) The avalanche rate versus the magnetic shear. (d) The average avalanche amplitude versus the magnetic shear.}
\label{fig:avalchar}
\end{figure}

Transport events of various sizes, called avalanches, are observed in a particular regime of KSTAR plasmas without macroscopic MHD instabilities (see Methods~\ref{sec:avalregime} and \ref{sec:kstaravst}).
The propagation of avalanche heat bumps is shown to be ballistic based on its propagation time ($\Delta t$) measurements over different distances ($\Delta R$). 
$T_\mathrm{e}$ time traces measured at two distant channels by $\Delta R$ are used to calculate the time lag for the maximum cross correlation which is taken as the propagation time for that distance.  
The reference channel ($\Delta R = 0$) is chosen as the one a few centimeters away from the avalanche initiation location $R_\mathrm{av}$ to minimize the possible interference of the initial avalanche propagation by the $E \times B$ staircase (see below). 
$\Delta R$ and $\Delta t$ have almost a linear relation $\Delta R \sim (\Delta t)^b$ where $b \approx 1$ as shown in Fig.~\ref{fig:avalchar}(a), indicating the ballistic propagation with the speed of $60.5 \pm 24.5$~m/s.
The measurement uncertainties in $\Delta R$ and $\Delta t$ represent the estimated uncertainty in the ECE channel position and the standard deviation of time lag measurements for a given period.
The uncertainty in the avalanche speed is not small because the linearity between $\Delta R$ and $\Delta t$ is not perfect, especially in the edge region. 
Nonetheless, it is clear that the overall propagation behavior of avalanche heat pulses is far from diffusive and close to ballistic, and this behavior is consistently observed. %in the most avalanche periods. 
Note that in the same plasma the approximated mean diamagnetic velocity in the bump propagation region is $ \overline{ \rho_s C_s / a } \approx 720$~m/s where the mean ion sound Larmor radius is $\overline{\rho_s} \approx 1.61 \times 10^{-3}$~m and the mean ion sound speed is $\overline{C_s} \approx 2.02 \times 10^5$~m/s. %
%More complicated behavior~\cite{GarbetPoP2007} is observed in the avalanche plasmas with the continuous electron cyclotron resonance heating, depending on its deposition location against $R_\mathrm{av}$.
%This will be reported in future work with more experiments. 

\subsection{The parametric dependency of the avalanche characteristics \label{sec:avalpara}}

The characteristics of avalanches such as the propagation speed, the avalanche rate (the number of avalanches in one second) and the avalanche amplitude (the prominence of the avalanche peak in $T_\mathrm{e}$) are measured, and their parametric dependency is investigated. 
Note that only sufficiently large avalanches are counted to minimize the noise contribution.
Measured avalanche characteristics are the representative values of 49--59 quasi-stationary periods of the multiple KSTAR avalanche plasmas (see Methods~\ref{sec:avalregime}). 
Each quasi-stationary period is about 0.2 sec long.  
Time averaged values of various physical parameters (the magnetic shear $\hat{s}=\frac{r}{q}\frac{dq}{dr}$ where $q$ is the safety factor, $R/L_{T_\mathrm{i,e}}$, $R/L_{n}$, the ion collisionality~\cite{SauterPoP1999}, etc) for each period are calculated and compared with the avalanche characteristics. 
The physical parameters of the periods differ by the natural plasma evolution (see Methods~\ref{sec:avalregime}) and by the different toroidal fields. 

Among various correlations or anti-correlations, some noticeable results are introduced in Figs.~\ref{fig:avalchar}(b), \ref{fig:avalchar}(c) and \ref{fig:avalchar}(d). 
Firstly, the avalanche propagation speed seems to depend on the stability parameter ($\eta_i \equiv L_n/L_{T_\mathrm{i}}$) of the ITG mode which is expected to be dominant in this plasma (see Methods~\ref{sec:avalregime}).
Assuming that the linear growth rate of the ITG follows $\gamma \sim \eta_i$, the scattered correlation in Fig.~\ref{fig:avalchar}(b) is consistent with the expected linear dependence of the turbulence front propagation speed on the linear growth rate in the theory~\cite{GurcanPoP2005}.
However, measurements are scattered significantly with large uncertainties, and there may be room for different interpretations. 
Secondly, as the magnetic shear is decreased, the avalanche seems to occur more frequently with a larger average amplitude.   
This might be related to the larger radial size of the ITG turbulence eigenmode with the lower magnetic shear~\cite{PearlsteinPRL1969, Miyamoto2007}, while the linear toroidal mode coupling would be less efficient due to the increased distance between rational surfaces~\cite{GarbetNF1994, Miyamoto2007}. %

Although the correlation can provide some hints, it should be taken carefully.
One reason is that plasma parameters are often strongly correlated with other plasma parameters in these experiments. 
The other is that averaged characteristics may not be a meaningful quantity to assess complex phenomena that have long tail distributions of characteristics.

\subsection{Edge turbulence eddies with avalanches \label{sec:avaleddy}}

\begin{figure}[t]
\includegraphics[keepaspectratio,width=0.6\textwidth]{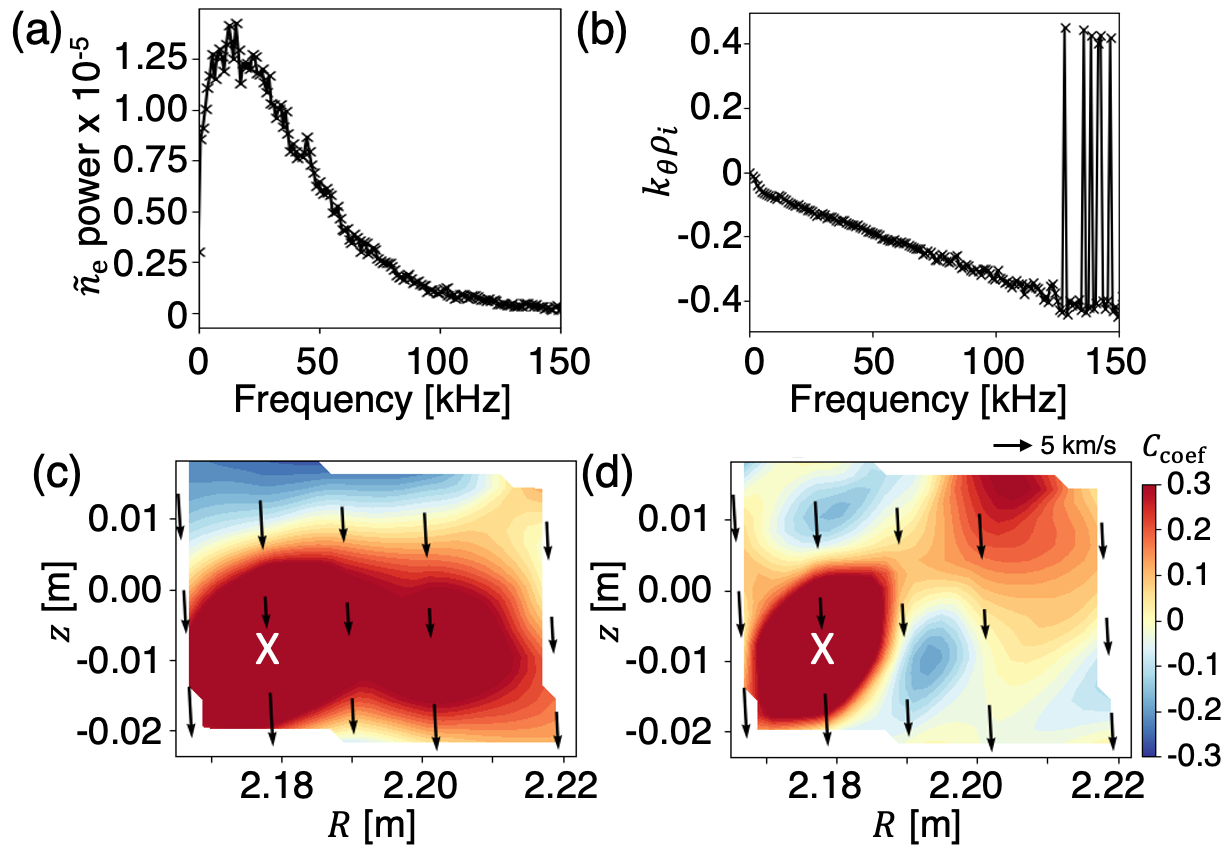}
\caption{(a) The normalized density fluctuation ($\tilde{n}_\mathrm{e} = \delta n_\mathrm{e} / \langle n_\mathrm{e} \rangle $) power spectrum. (b) The local wavenumber frequency spectrum. The cross correlation coefficient ($C_\mathrm{coef}$) images of density fluctuation (50--100 kHz) at the zero time lag in (c) the near-large-avalanche phases and (d) the far-from-large-avalanche phases. The white X indicates the reference channel for the $C_\mathrm{coef}$ calculation. The length of the black arrows means the strength of the local laboratory phase velocity $v^\mathrm{ph}_\mathrm{lab}$. }
\label{fig:avaleddy}
\end{figure}

Edge turbulence eddies in the KSTAR avalanche plasmas are identified as the ion modes, rotating in the ion diamagnetic drift direction. 
It is found using the two-dimensional edge density fluctuation measurements from the beam emission spectroscopy (BES) diagnostics~\cite{Nam:2014kd} in the edge region of $R = 2.166$--$2.218$~m near the midplane (see Methods~\ref{sec:diagmeth}).
The density fluctuations from two poloidally adjacent channels are used to calculate the cross power spectrum and the local wavenumber frequency spectrum shown in Figs.~\ref{fig:avaleddy}(a) and \ref{fig:avaleddy}(b), respectively. 
The significant fluctuations are detected in the frequency range of 0--120~kHz and the wavenumber range of $-0.4 \le k_\theta \rho_i \le 0$. 
This means a negative (clockwise) poloidal phase velocity in the laboratory frame, i.e. $v^\mathrm{ph}_\mathrm{lab} = v^\mathrm{ph}_\mathrm{plasma} + v_{E \times B} = -3.51 \pm 0.41$~km/s. 
Since the equilibrium $E \times B$ velocity is found to be positive, the phase velocity in the plasma frame should be negative along the ion diamagnetic drift direction.

Turbulence eddies are found to form a radially extended coherent structure in the near-large-avalanche phases. 
Here, the large avalanche means an avalanche whose bump is sufficiently large to propagate to the edge region.  
The structure of turbulence eddies is investigated using the cross correlation coefficient ($C_\mathrm{coef}$) image between the reference channel (whose location is marked by the white X) and other channels at the zero time lag. 
The set of the 991 density fluctuation (50--100~kHz) measurements in the near-large-avalanche phases is used for Fig.~\ref{fig:avaleddy}(c), and the set of the 990 measurements in the far-from-large-avalanche phases (in between the large avalanches) is used for Fig.~\ref{fig:avaleddy}(d). 
The extended coherent structure of turbulence eddies is observed in Fig.~\ref{fig:avaleddy}(c).
Moreover, it is found to appear and disappear quasi periodically with respect to the large avalanche, implying the close relation between this kind of extended structure~\cite{SarazinNF2010, KinPoP2019, KishimotoPTRA2023, HongPoP2023} by the mode-mode interaction and the large avalanche (successive interaction)~\cite{DiamondPoP1995, HahmJKPS2018}. 

The arrows in Figs.~\ref{fig:avaleddy}(c) and \ref{fig:avaleddy}(d) indicate the local (near poloidal) $v^\mathrm{ph}_\mathrm{lab}$ measurements using the cross phase of the 50--100 kHz density fluctuations between poloidally adjacent BES channels. 
The $v^\mathrm{ph}_\mathrm{lab}$ measurements in Figs.~\ref{fig:avaleddy}(c) and \ref{fig:avaleddy}(d) are more or less similar.

\subsection{The avalanche pseudo-size distribution \label{sec:avaldist}}

\begin{figure}[t]
\includegraphics[keepaspectratio,width=0.6\textwidth]{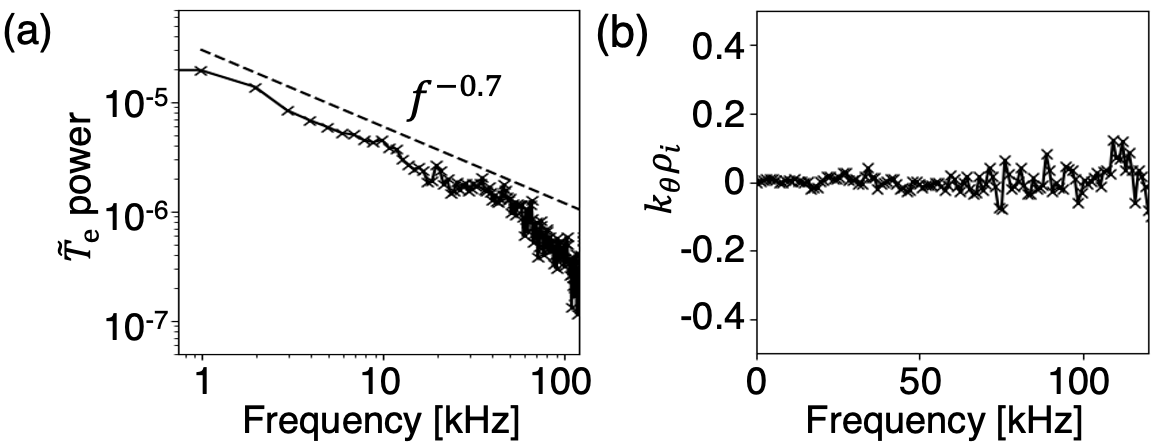}
\caption{(a) The normalized electron temperature fluctuation power spectrum. (b) The local wavenumber frequency spectrum.}
\label{fig:avalsize}
\end{figure}

In the KSTAR avalanche plasmas, avalanches of various sizes and scales are observed (see Methods~\ref{sec:kstaravst})~\cite{Choi:2019wy}.
In the previous analysis~\cite{Choi:2019wy}, the power-law correlation function (the power-law frequency spectrum of $|\delta T_\mathrm{e} / \langle T_\mathrm{e} \rangle|^2$, $S(f) \propto f^{-0.7}$) was shown with the large Hurst exponent $H > 0.5$, suggesting the long range temporal scale or the temporal criticality of events~\cite{SanchezPPCF2015}. 
On the other hand, the frequency spectrum of $|\delta T_\mathrm{e} / \langle T_\mathrm{e} \rangle|^2$ might be also thought of as the pseudo-size distribution of avalanches.
The size in the context of the SOC avalanche~\cite{Jensen1998} is defined as the volume of the temperature perturbation by one avalanche, i.e. $S_\mathrm{av} = \mathrm{height} \times \mathrm{width} \times \mathrm{length} = \delta T \times \delta R \times 2\pi Rq$. 
Note that it can be related to a heat flux carried by each avalanche, $Q \sim \frac{3}{2} n \delta T \delta v_r$~\cite{KinNF2023} where $\delta v_r$ is the avalanche propagation speed, assuming that $\delta R$ is associated with $\delta v_r$.
It can be further written that $S_\mathrm{av} \sim \delta T \delta R \sim \delta T \delta v_r \sim \delta T^2 \sim S$ with $\delta v_r \sim \delta T$~\cite{Kosuga:2013io}. 
The $|\delta T_\mathrm{e} / \langle T_\mathrm{e} \rangle|^2$ frequency spectrum shown in Fig.~\ref{fig:avalsize}(a) follows the power-law $S(f) \propto f^{-0.7}$, meaning that the probability distribution of $S$ also follows the power-law $P(S) \sim S^{-1/0.7}$ which can be considered the distribution of the avalanche pseudo-size. 
The frequency spectrum is obtained by the cross power between two poloidally adjacent measurements using the ECE imaging channels near $R_\mathrm{av}$~\cite{Choi:2019wy}.
The local wavenumber frequency spectrum in Fig.~\ref{fig:avalsize}(b) shows that the measured temperature fluctuations correspond to temperature variations by transport process ($m=0$) rather than turbulence eddies. 

\subsection{The $E \times B$ staircase and its avalanche stopping capability \label{sec:stairstop}}

\begin{figure}[t]
\includegraphics[keepaspectratio,width=0.6\textwidth]{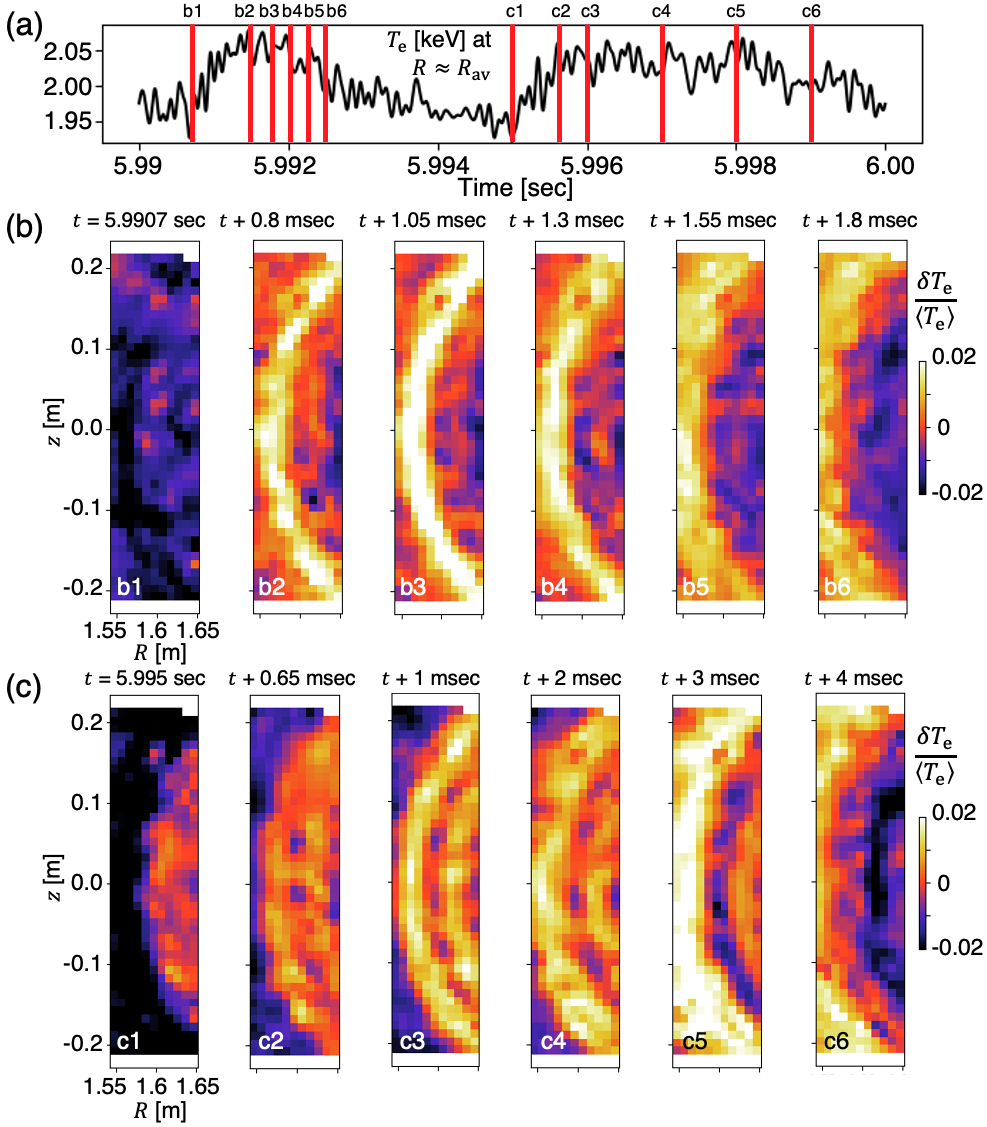}
\caption{(a) The $T_\mathrm{e}$ time trace at $R \approx R_\mathrm{av}$. Measurement times of $\delta T_\mathrm{e} / \langle T_\mathrm{e} \rangle$ images shown in (b) and (c) are indicated by red lines and letters ‘b’s and ‘c’s, respectively. $\delta T_\mathrm{e} / \langle T_\mathrm{e} \rangle$ images along time ($t$) during the large avalanche (b) without and (c) with the $E \times B$ staircase.}
\label{fig:stairaval}
\end{figure}

The $E \times B$ staircase manifests as the temperature profile corrugation, i.e. the multiple transport barriers confining avalanches within the mesoscale tread width (see Methods~\ref{sec:kstaravst}).
The comparison of propagations of the large avalanche without and with the staircase demonstrates its avalanche stopping capability~\cite{GhendrihEPJD2014} as shown in Fig.~\ref{fig:stairaval}. 
Figs.~\ref{fig:stairaval}(b) and \ref{fig:stairaval}(c) show $\delta T_\mathrm{e} / \langle T_\mathrm{e} \rangle$ images during propagations of bumps and voids of the large avalanche without and with the staircase, respectively. 
They are observed consecutively, and their bumps detected in the local $T_\mathrm{e}$ measurement near $R \approx R_\mathrm{av}$ are shown in Fig.~\ref{fig:stairaval}(a) with the indication of the times of the images (red lines). 
In Fig.~\ref{fig:stairaval}(b) without the staircase it takes about 1.8~msec for the local bump ($\delta T_\mathrm{e} / \langle T_\mathrm{e} \rangle > 0$) to escape the diagnostics view, while in Fig.~\ref{fig:stairaval}(c) with the staircase (jet-like $\delta T_\mathrm{e} / \langle T_\mathrm{e} \rangle$ patterns) it takes about 4~msec. 
The staircase impedes the propagation of the avalanche heat flux significantly, acting as transport barriers for a moment. 
 
On the other hand, Fig.~\ref{fig:stairaval}(c) can be the direct observation of the dynamics between avalanches and the $E \times B$ staircase, phenomenologically consistent with one idea about their relation, i.e. the jam instability~\cite{Kosuga:2013io}. 
With the time delay between variations of a field and its flux, the avalanche flux propagation can be jammed to form a seed perturbation (temperature corrugation) of the transport barrier~\cite{Kosuga:2013io}. 
Note that recent gyrokinetic simulations found that there exists such a time delay in both temperature~\cite{MutoPoP2021} and density~\cite{QiNF2022}.
The jamming occurs when an avalanche is so fast to overtake a wave that can be carried by the system. 
Since the avalanche speed depends on the avalanche heat pulse size in the model~\cite{Kosuga:2013io}, it results in the condition that the heat pulse size should be larger than some threshold value for the jamming.
The relation between avalanches and the staircase will be further discussed later (see below).

\subsection{The initial formation and transformation of the $E \times B$ staircase \label{sec:stairdyn}}

\begin{figure}
\includegraphics[keepaspectratio,width=0.7\textwidth]{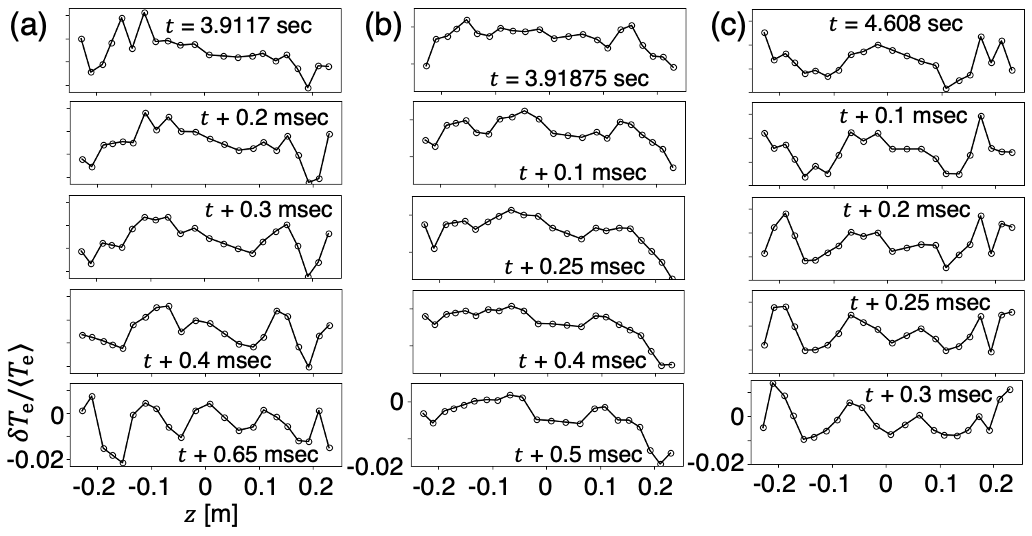}
\caption{(a) The initial formation of the $E \times B$ staircase (class A). (b) Transformation of the smaller scale ($\Delta_z \approx 10$~cm) staircase to the larger scale ($\Delta_z \approx 20$~cm) staircase. (c) Transformation of the larger scale ($\Delta_z \approx 25$~cm) staircase to the smaller scale ($\Delta_z \approx 12$~cm) staircase.}
\label{fig:stairdyn}
\end{figure}

The initial formation and transformations of the $E \times B$ staircase have been revealed by investigating the long range vertical $\delta T_\mathrm{e} / \langle T_\mathrm{e} \rangle$ profiles obtained in the previous experiment (see Methods~\ref{sec:kstaravst})~\cite{Choi:2019wy}. 
Fig.~\ref{fig:stairdyn}(a) shows the formation process of the $E \times B$ staircase. 
Initially, a localized perturbation near $R_\mathrm{av}$ arises.  
The perturbation grows in size, propagates toward the plasma center with the speed of 100 m/s which is anomalously fast~\cite{LeePRL2003}, and results in the globally self-organized mini barriers with the mesoscale vertical tread width $\Delta_z = 10.8$~cm. 
They dissipate in time with the appearance of an $m=1$ structure for about 1~msec, but reform repeatedly~\cite{Choi:2019wy}. 
In addition, the change of the $E \times B$ staircase width is observed as shown in Figs.~\ref{fig:stairdyn}(b) and \ref{fig:stairdyn}(c). 
In Fig.~\ref{fig:stairdyn}(b) the weak amplitude and small width staircase ($\Delta_z \approx 10$~cm) grows to have a stronger amplitude and larger width ($\Delta_z \approx 20$~cm).
It is worth noting that the merging of small scale transport barriers~\cite{HughesAJ2021} has been considered one possible way to form a large scale strong transport barrier~\cite{AshourvanPRE2016}. 
For example, a sudden merger or disappearance of multiple zonal-band structures leading to a few broader zonal jets has been observed in long term simulation of forced two-dimensional barotropic incompressible flows in the geophysical fluid dynamics context~\cite{ObusePF2010}.
%Although the resulting larger barriers in Fig.~\ref{fig:stairdyn}(b) dissipate in time possibly through the collisional damping, the increased $\delta T_\mathrm{e}$ amplitude implies the stronger transport regulation for the larger width staircase. 
In Fig.~\ref{fig:stairdyn}(c) the opposite transformation is shown. 
The initial $\Delta_z \approx 25$~cm staircase becomes about the half width within 0.3~msec. 
Observations in KSTAR plasmas show that the $E \times B$ staircase is persistent but not very stationary (destruction, reformation and transformation), which is consistent with the observations in the gyrokinetic simulations~\cite{GhendrihEPJD2014, DifNF2017}.

\subsection{Edge turbulence eddies with the $E \times B$ staircase \label{sec:staireddy}}

\begin{figure}
\includegraphics[keepaspectratio,width=0.6\textwidth]{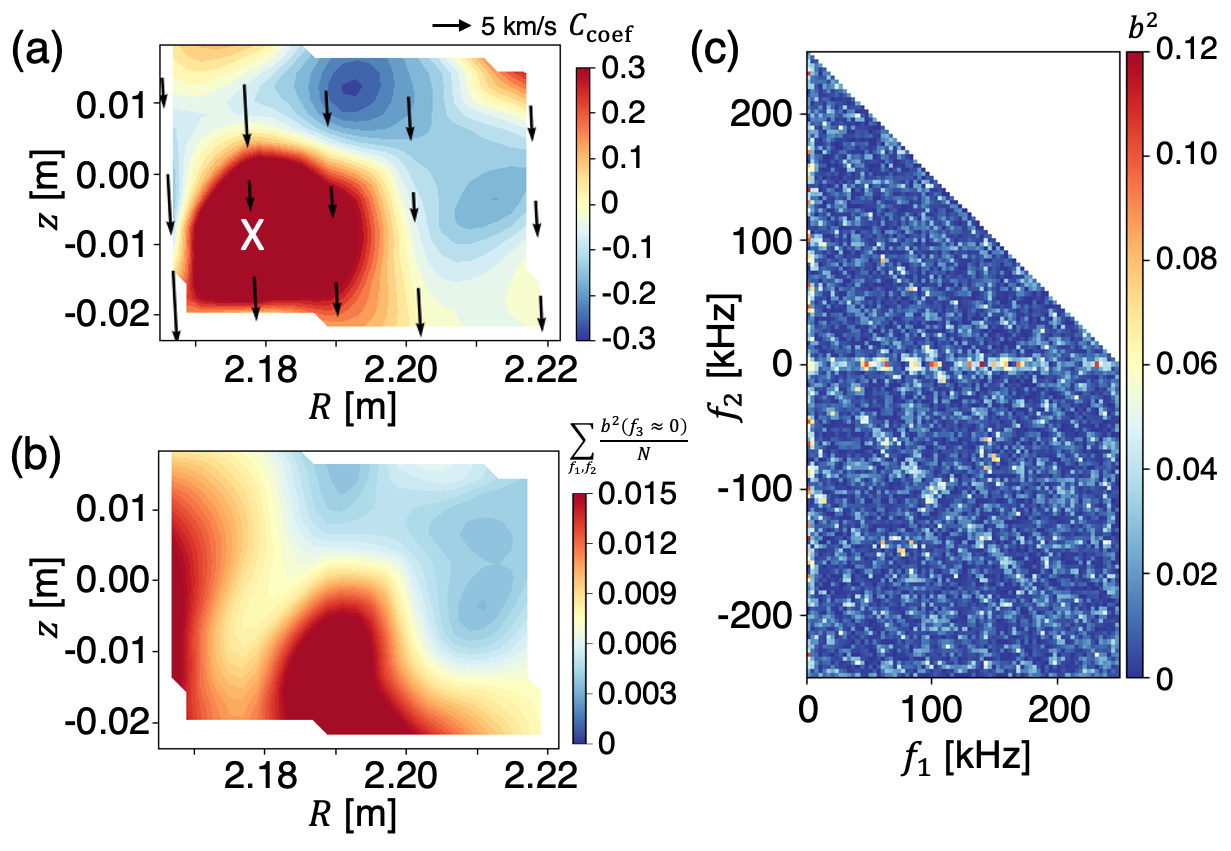}
\caption{(a) The cross correlation coefficient ($C_\mathrm{coef}$) images of density fluctuation (50--100 kHz) at the zero time lag with the $E \times B$ staircase. The white X indicates the reference channel for the $C_\mathrm{coef}$ calculation. The length of the black arrows means the strength of the local laboratory phase velocity $v^\mathrm{ph}_\mathrm{lab}$. (b) The average bicoherence image with $f_3 \approx 0$~kHz. (c) The bicoherence of density fluctuations at $R=2.19$~m and $z=-0.02$~m. }
\label{fig:staireddy}
\end{figure}

The structure of edge turbulence eddies during the $E \times B$ staircase phases is found to be radially more limited, compared to that in the near-large-avalanche phases in Fig.~\ref{fig:avaleddy}(c). 
In Fig.~\ref{fig:staireddy}(a), the cross correlation coefficient ($C_\mathrm{coef}$) image at the zero time lag is shown. 
It is obtained using the set of the 102 density fluctuation (50--100~kHz) measurements in the staircase phases during which the staircase of the radial tread width $\Delta =2$--$3$~cm exists. 
It seems that the radial extent of the strongly correlated region is limited by the tread width of the $E \times B$ staircase, i.e. the distance between the shear flow layers.

The analysis of the auto-bicoherence in density fluctuations suggests that the zonal density ($\sim$ potential) field, constituting the $E \times B$ shear flow layer, is enhanced near the edge of the correlated structure through the nonlinear energy transfer from turbulence. 
Fig.~\ref{fig:staireddy}(b) shows the two-dimensional plot of the average local bicoherence with $f_3 \approx 0$. 
The bicoherence measures the degree of the three-wave coupling among fluctuations of frequencies $f_1$, $f_2$, and $f_3 = f_1 + f_2$~\cite{Kim:1979ps, Choi:2019tw}.  
The local density fluctuation measured by each BES channel is used to calculate the local squared auto-bicoherence ($b^2$) shown in Fig.~\ref{fig:staireddy}(c).
The $b^2$ is averaged over the fluctuation pairs with $f_3 \approx 0$ (along the $f_2 = - f_1$ line) to give $\sum_{f_1,f_2} b^2(f_3 \approx 0)/N$ where $N$ is the number of fluctuation pairs with $f_3 \approx 0$.
The relatively large $\sum_{f_1,f_2} b^2(f_3 \approx 0)/N$ is observed near both edges ($R=2.167$~m and $R=2.19$~m) of the correlated structure, implying that the developed zonal flow shear there interferes with the spatial coupling of turbulence eddies.  
Note that the local $v^\mathrm{ph}_\mathrm{lab}$ measurements in the staircase phases, indicated by the black arrows in Fig.~\ref{fig:staireddy}(a), show the significant variation around the large $\sum_{f_1,f_2} b^2(f_3=0)/N$ regions (for example, at $R=2.167$~m and $z=-0.015$~m), compared to the $v^\mathrm{ph}_\mathrm{lab}$ measurements in the near-large-avalanche phases in Fig.~\ref{fig:avaleddy}(c).

%It seems that the length of the arrow (the strength of the $v^\mathrm{ph}_\mathrm{lab}$) increases on the right side of the large $\sum_{f_1,f_2} b^2(f_3=0)/N$ region (for example, at $R=2.167$~m and $z=-0.015$~m) and decreases on the left side (for example, at $R=2.178$~m and $z=-0.015$~m).
%This implies the positive zonal potential perturbation in the large $\sum_{f_1,f_2} b^2(f_3\approx0)/N$ region. 

\subsection{The parametric dependency of the $E \times B$ staircase characteristics \label{sec:stairpara}}

The characteristics of the $E \times B$ staircase such as the staircase ratio, the lifetime, the power and the radial tread width are measured, and their parametric dependency is investigated. 
The Lomb-Scargle periodograms~\cite{VanderPlas2018} of non-uniform $\delta T_\mathrm{e} / \langle T_\mathrm{e} \rangle$ measurements and the cluster algorithm~\cite{OPTICS, SKLEARN} are used to identify the staircase and extract its characteristics (see Methods~\ref{sec:diagmeth}). 
The measurements in the aforementioned quasi-stationary periods are used, and the same caution would apply. 

\begin{figure}
\includegraphics[keepaspectratio,width=0.6\textwidth]{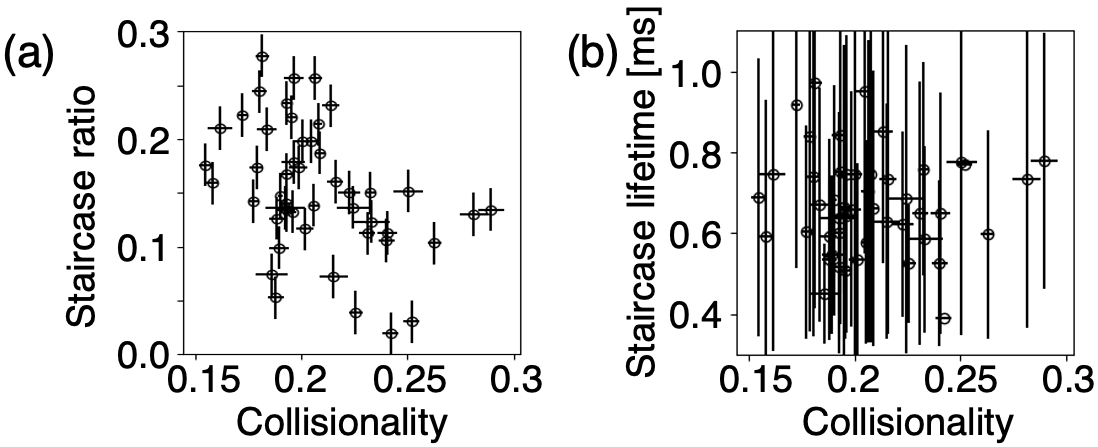}
\caption{(a) The $E \times B$ staircase ratio versus the ion collisionality. (b) The average $E \times B$ staircase lifetime versus the ion collisionality.}
\label{fig:staircoln}
\end{figure}

The most noticeable relation is found between the staircase ratio and the ion collisionality~\cite{SauterPoP1999} as shown in Fig.~\ref{fig:staircoln}(a). 
The staircase ratio means the staircase-detected-fraction of each analysis period. 
The overall anti-correlation between the staircase ratio and the collisionality means that the staircase is observed for a relatively shorter time in the higher collisionality regime. 
The average lifetime and the collisionality in Fig.~\ref{fig:staircoln}(b) have shown the more scattered anti-correlation.
Note that the lifetime distribution seems to have a long tail (see below). 

\subsection{Distributions of the $E \times B$ staircase characteristics \label{sec:stairdist}}

Numerous measurements of the $E \times B$ staircase characteristics were obtained for the similar 23 quasi-stationary periods from 4 repeated discharges to investigate their distributions. 
These similar quasi-stationary periods are characterized by the magnetic shear $\hat{s}=\frac{r}{q}\frac{dq}{dr}=0.23$--$0.35$, the normalized ion and electron temperature gradients $R/L_{T_\mathrm{i}} =7.5$--$9$ and $R/L_{T_\mathrm{e}} = 9.0$--$10.5$, the ion collisionality~\cite{SauterPoP1999} $= 0.18$--$0.215$ and plasma stored energy $= 300$--$320$~kJ. 

\begin{figure}[t]
\includegraphics[keepaspectratio,width=0.8\textwidth]{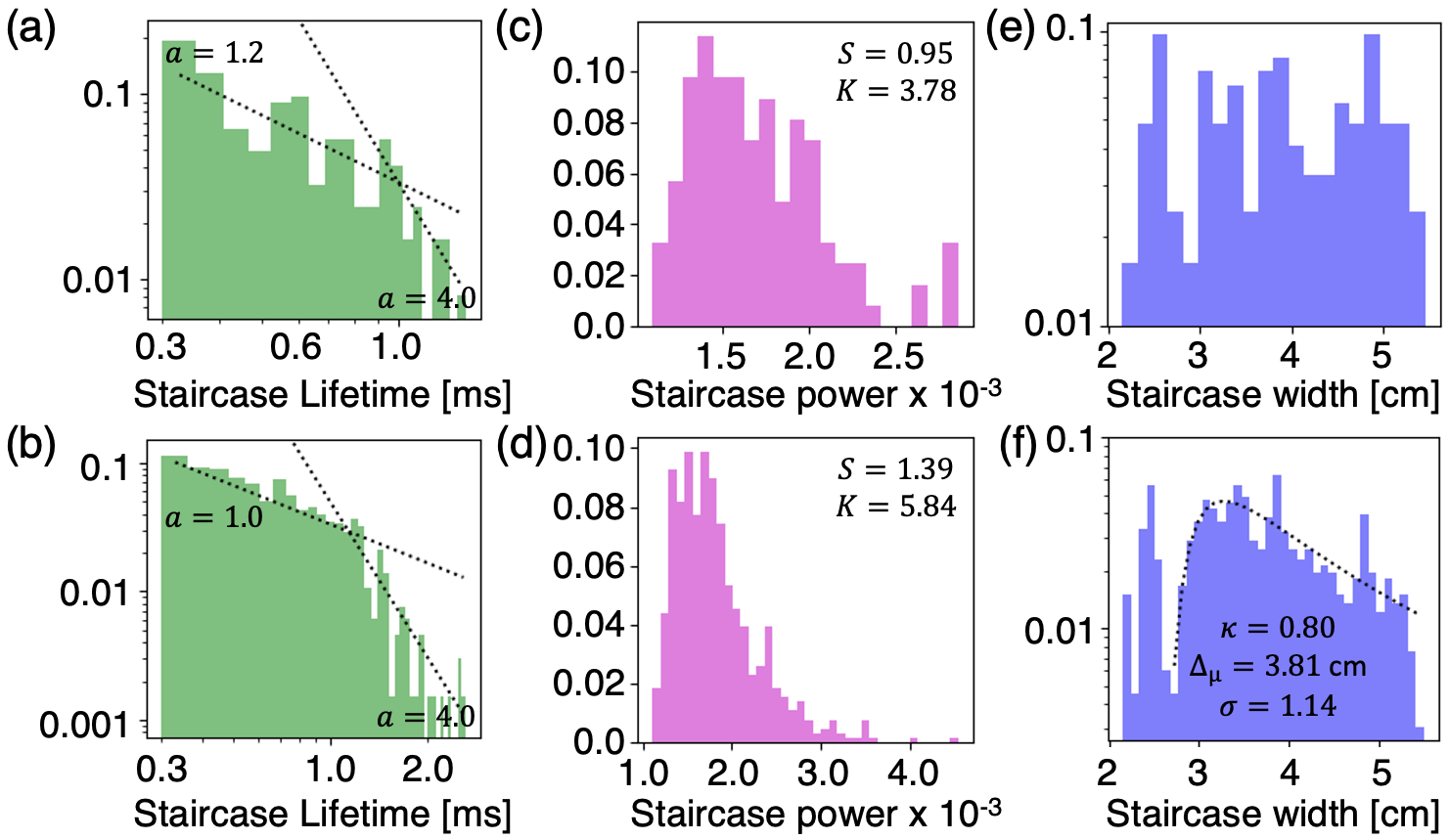}
\caption{Normalized probability distributions of the lifetime, the power, and the radial tread width of the $E \times B$ staircase (a, c, e; top) away from the large avalanche (class A) and (b, d, f; bottom) near the large avalanche (class B), respectively. Dotted lines indicate constant multiples of power-law $t_\mathrm{life}^{-a}$ for the lifetime distribution and a Fr\'echet distribution with the shaping ($\kappa$), location ($\Delta_\mu$) and scale ($\sigma$) parameters for the width distribution.}
\label{fig:stairstat}
\end{figure}

As explained in Methods~\ref{sec:diagmeth}, the $E \times B$ staircase is classified into two classes A (away from the large avalanche) and B (near the large avalanche) depending on its proximity to the large avalanche. 
A total of 123 staircases are found for the class A and 659 staircases for the class B. 
The lifetime ($t_\mathrm{life}$), the power ($|\delta T_\mathrm{e} / \langle T_\mathrm{e} \rangle|^2$) and the radial tread width ($\Delta$) probability distributions of the class A barriers are shown in Figs.~\ref{fig:stairstat}(a), \ref{fig:stairstat}(c), and \ref{fig:stairstat}(e), and those of the class B are shown in Figs.~\ref{fig:stairstat}(b), \ref{fig:stairstat}(d) and \ref{fig:stairstat}(f), respectively. 

Distributions of the lifetime and power do not differ much between the class A and the class B.
For example, the lifetime distributions of both classes can be fitted with a power-law ($t_\mathrm{life}^{-a}$) function with a coefficient $a$ changing around 1~msec as shown in Figs.~\ref{fig:stairstat}(a) and \ref{fig:stairstat}(b). 
The power-law distribution of the lifetime implies the non-existence of the characteristic lifetime and the critical behavior in time.
However, the range of the lifetime may be too narrow to conclude, and an exponential distribution ($e^{-a t_\mathrm{life}}$) with a characteristic time ($1/a$) also provided a fair agreement. 
The power probability distributions of both classes have shown a positive skewness $S > 0$ and a kurtosis $K > 3.0$ as shown in Figs.~\ref{fig:stairstat}(c) and \ref{fig:stairstat}(d). 
The non-Gaussianity of distributions means that the large power staircase is intermittently observed. 

Unlike other distributions, the radial tread width ($\Delta$) probability distributions of the class A and the class B look different. 
While the distribution of the class A seems more or less random, the distribution of the class B has a positive skewness fat tail for $\Delta \ge 2.7$ cm with an independent peak around $\Delta \approx 2.5$ cm. 
It is found that the fat tail of the $\Delta$ distribution of the class B shows reasonable agreement with the constant multiple of a Fr\'echet distribution $F(\Delta) = \frac{\tau^{1+\kappa}}{\sigma} e^{-\tau}$ with $\tau=(1 + \kappa \frac{\Delta - \Delta_\mu}{\sigma})^{-1/\kappa}$, where the shaping parameter $\kappa$ = 0.80, the location parameter $\Delta_\mu$ = 3.81 cm and the scale parameter $\sigma$ = 1.14.
The parameters are obtained by the least-square fit.
%By the Anderson-Darling test~\cite{StephensJASA1974, ScholzJASA1987}, the null hypothesis that the class B width samples follow the Fr\'echet distribution with the given parameters may not be rejected since the p-value is found to be higher than 25 \% (the test is capped at 25 \%). 
This is qualitatively consistent with the gyrokinetic simulation result~\cite{DifNF2017}, where the width measurements were obtained over a broad range of normalized gyro-radius, and their distribution follows a Fr\'echet distribution with different parameters~\cite{MilovanovPRE2021}. 
The distinguished Fr\'echet distribution for the $E \times B$ staircase close to the large avalanche might result from the influence of the large avalanche on the staircase formation as discussed below.
Note that, when not classified, the width distribution of all staircases also follows well a Fr\'echet distribution.
The proximity to the large avalanche is the particular parameter among others which makes a clearest distinction in width distributions by classification.
For the ion collisionality, the relatively lower collisionality class seems to have a shifted distribution toward the higher width in a similar fashion of zonal jets evolution in geophysical fluid~\cite{ManfroiJAS1999}, but more data over the wider range of the collisionality are required to conclude.

\section{Discussion \label{sec:disc}}

%The $E \times B$ staircase, globally self-organized mini transport barriers, is observed in the KSTAR avalanche transport regime~\cite{Choi:2019wy}.
%The detail experimental approach and methods for the detection and analyses are explained in Section~\ref{sec:kstaravst}. 
%Since the characteristics of underlying transport is closely associated with the mechanism and characteristics of the self-organization phenomena in the non-equilibrium complex system, various aspects of both avalanches and the $E \times B$ staircase have been analyzed and presented in this paper.
Various aspects of both avalanches and the $E \times B$ staircase have been analyzed and presented in this paper.
Before discussing how these analyses can lead to an improved understanding of their relation, the main findings are summarized in Table~\ref{tab:summary}.

\begin{table}
\includegraphics[keepaspectratio,width=\textwidth]{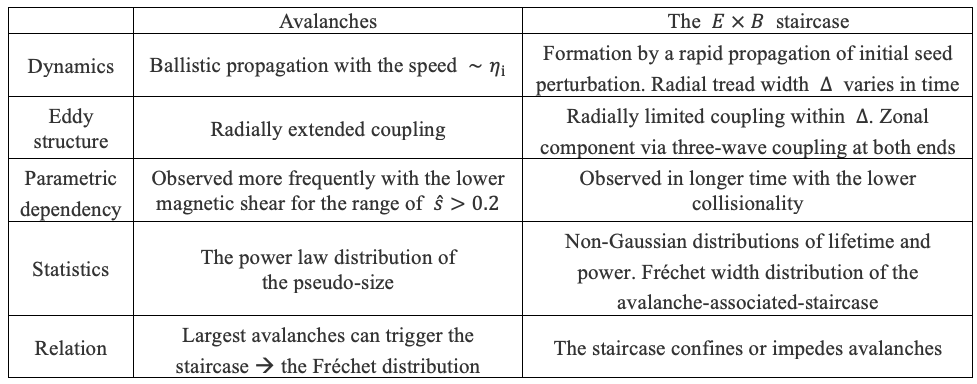}
\caption{Summary of avalanches and the $E \times B$ staircase analyses in KSTAR plasmas. }
\label{tab:summary}
\end{table}

To begin the discussion, the higher likelihood of finding the $E \times B$ staircase near the large avalanche suggests that the staircase requires an initial heat and/or momentum perturbation larger than some threshold value.
This naturally relates avalanches with the staircase and also explains the Fr\'echet-like width distribution for the staircase near the large avalanche.   
The required perturbation for the staircase can be provided by the large avalanche. 
The power-law distribution of the avalanche pseudo-size $P(S) \sim S^{-\alpha}$ means that the large event, which can satisfy the threshold, is rare but occurs.
The size of the heat perturbation may not be a sufficient condition, because there is the large avalanche having a comparable heat perturbation but not followed by the staircase formation (Fig.~\ref{fig:stairaval}(b)).
As pointed out in the gyrokinetic simulation~\cite{WangNF2018}, the radial synchronization and the synergetic effect between the mean $E_\mathrm{r}$ field variation by the large avalanche and the Reynolds zonal field can be important. 
Note that both the threshold condition and the positive feedback are important requisites for many self-organization phenomena in the non-equilibrium complex system.
On the other hand, the Fr\'echet distribution can follow if it is assumed that the staircase width ($\Delta$) depends on the size of the associated large avalanche (with $S_\mathrm{av} \sim \delta T \delta R$)~\cite{WangNF2018}. 
The Fr\'echet distribution is one of the extreme value distributions, corresponding to a distribution of maxima of samples when samples follow a distribution of the long tail such as the power-law function~\cite{Rinne2008}. 
Since the avalanche pseudo-size follows the power-law distribution $P(S) \sim S^{-\alpha}$ and the staircase threshold can be satisfied with the rare and large avalanche (the maximum of each long period, $\max{\{S\}}$), the staircase width distribution $P(\Delta \sim \max \{S\})$ would follow the Fr\'echet distribution.

\begin{figure}
\includegraphics[keepaspectratio,width=0.6\textwidth]{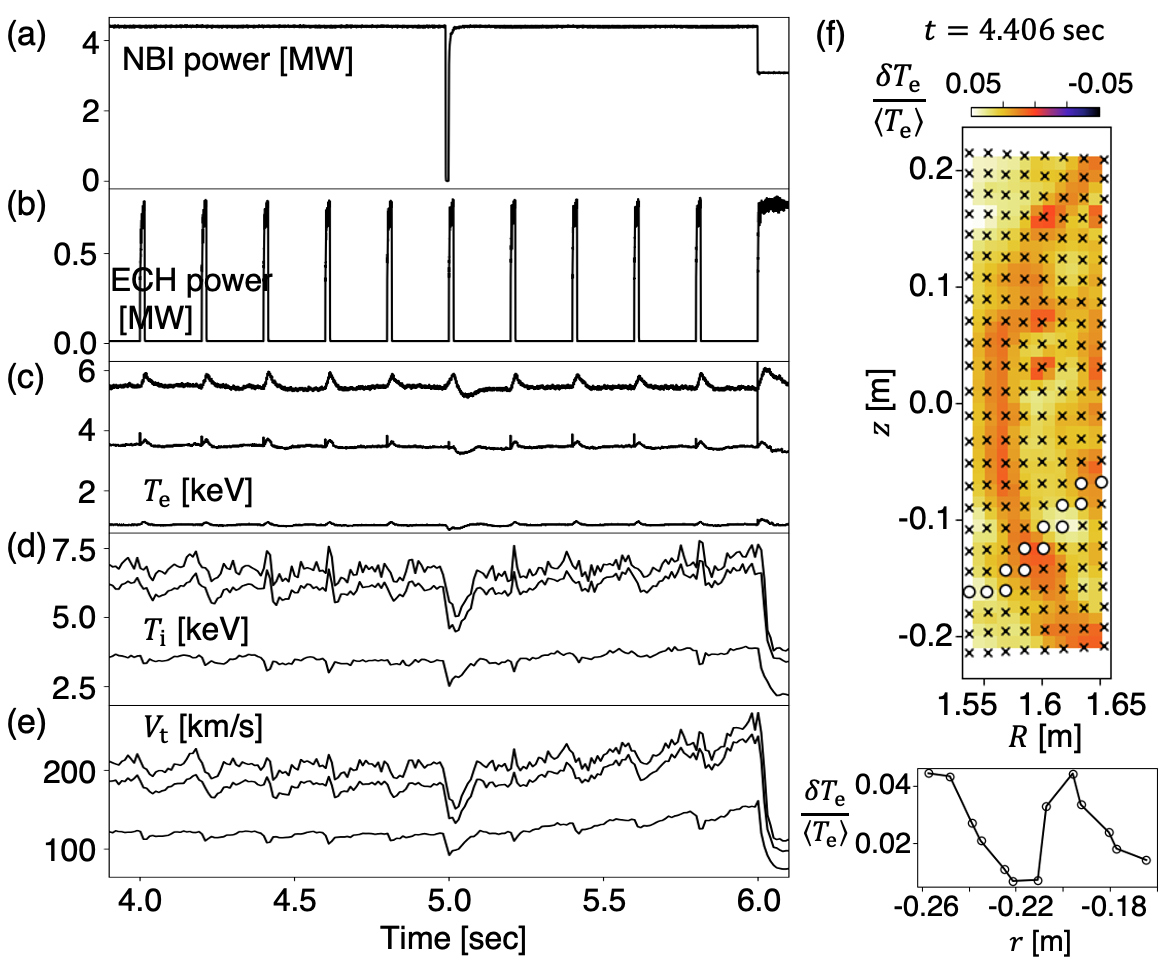}
\caption{(a) Total NBI power. (b) ECH power. (c, d, e) Electron temperature, ion temperature and ion toroidal velocity measured at different radial locations ($r/a \sim 0, 0.2, 0.6$ for $T_\mathrm{e}$ and $r/a \sim 0, 0.2, 0.4$ for $T_\mathrm{i}$ and $V_\mathrm{t}$), respectively. (f) The radial $\delta T_\mathrm{e} / \langle T_\mathrm{e} \rangle$ profile and the $\delta T_\mathrm{e} / \langle T_\mathrm{e} \rangle$ images at 4.406~sec. }
\label{fig:stairmech}
\end{figure}

For validation of the staircase threshold, controlled perturbation and response experiments would be required.
Although the fine scan of the perturbation size was not possible, the constant external heat perturbation could be applied using the modulated electron cyclotron resonance heating (MECH) to investigate the response of the KSTAR avalanche plasma as shown in Fig.~\ref{fig:stairmech}. 
The short pulses of the MECH induced the rapid and significant perturbation on the electron and ion temperature as well as the ion momentum over the almost entire radial region. 
The size of the MECH perturbation on the electron temperature was almost twice that of the large avalanche observed around.  
Interestingly, the $E \times B$ staircase of the relatively large width $\Delta \ge 6$~cm is partly observed a few milliseconds after every MECH heat pulse injection. 
The one example at $4.406$~sec is shown in Fig.~\ref{fig:stairmech}(f).
This suggests that the assumption made in the previous discussion, i.e. the staircase width depends on the associated perturbation size, can be valid, though more quantitative experiment and analysis are necessary. 
Note that the jam instability model of the staircase~\cite{Kosuga:2013io}, introduced briefly in Section~\ref{sec:stairstop}, suggested the threshold condition on the heat pulse size, originated from the dependency of the avalanche propagation speed on the heat pulse size.  
The propagation speed of the MECH perturbation is found to be also relatively higher $\sim 100$~m/s, compared to the heat bump propagation speed of the large avalanche ($< 75$~m/s) in the same plasma. 

Secondly, both avalanches and the $E \times B$ staircase exhibit the critical features that their initial localized perturbation propagates rapidly over the whole system (Figs.~\ref{fig:avalchar}(a) and \ref{fig:stairdyn}(a)) and their distributions have a long tail like the power-law. 
The central task would be to understand how their critical features manifest and how they are related to the other non-equilibrium critical phenomena. 
The critical features of avalanches in tokamak plasmas have been relatively better understood in association with the self-organized criticality (SOC)~\cite{DiamondPoP1995}. 
The avalanche in the SOC system operates by the domino effect over the near marginal profile, i.e. the successive destabilization of nearby locations by the flux propagation~\cite{HahmJKPS2018, SanchezPPCF2015}. 
However, how the localized momentum perturbation propagates the long range and the correlated structure of zonal flows, the $E \times B$ staircase, appears are less understood. 
The inter-shear-layer avalanches might mediate their interactions by delivering the Reynolds stress momentum~\cite{MilovanovPRE2021, ZhuPRL2020} (Fig.~\ref{fig:staireddy}(b)).
The power-law-like distribution of the staircase lifetime (Figs.~\ref{fig:stairstat}(a) and \ref{fig:stairstat}(b)) and the fast propagation of the initial perturbation (Fig.~\ref{fig:stairdyn}(a)) might imply another kind of avalanche composed of the momentum flux propagation~\cite{KuNF2012, GurcanJPA2015} in the SOC system.
Or, the phase space structure in the near-marginal regime might play a role in the long range interaction between zonal flows~\cite{Diamond2010, KosugaPoP2011}.

%\begin{figure}[t]
%\includegraphics[keepaspectratio,width=1.0\textwidth]{figure13.png}
%\caption{Analogy between the directed percolation (DP) in the space-time ($x+t$) space and the staircase dynamics in the wavenumber-time ($k+t$) space.
%Diffusion (Figs.~\ref{fig:stairdyn}(b) and \ref{fig:stairdyn}(c)), offspring and coalescence of the staircase in the $k+t$ space were observed in the experiment.}
%\label{fig:stairdydp}
%\end{figure}

Nonetheless, considering the $E \times B$ staircase as the self-organization near a non-equilibrium critical state can provide a new perspective to understand its characteristics and evolution by an analogy with a directed percolation (DP). 
%The directed percolation is one of the most studied non-equilibrium critical phenomena~\cite{Livi2017}.
%Its typical configuration is the two-dimensional tilted lattice where each site is either wet or dry and connected by bonds.
%Each bond is active with the probability $p$, and the wet site propagates in a given direction through the active bond.
%The parallel axis to the propagation is taken as the time ($t$) axis, and the spatio-temporal evolution of the system can be studied as $p$ varies. 
%If $p$ is smaller than some critical value $p_\mathrm{c}$, the system ends up with a fully dry absorbing state with finite correlation lengths.
%If $p \ge p_\mathrm{c}$, the system falls into an active state with infinitely percolating wet sites. 
The non-stationary and persistent dynamics of the staircase in the wavenumber-time ($k+t$) space (see Methods~\ref{sec:diagmeth}) can be thought of as a DP process with slow drive and many absorbing states.
The power-law-like distribution of the lifetime is not inconsistent with this view of the staircase (close to temporal criticality).
Then, near the active state (criticalities in both time and wavenumber space), the infinitely percolating modes ($k$s) in the staircase $k+t$ space would correspond to the one large long standing transport barrier, i.e. the conventional internal or edge transport barrier~\cite{ChungNF2017, Ashourvan:2019ek}. 

Finally, suggestions for future research will be mentioned briefly. 
Firstly, in this paper, most analyses are based on the electron temperature and density measurements due to the lack of a high resolution ion diagnostics.
Significant variations of the ion temperature and velocity by the largest avalanche in the KSTAR plasma were identified, but the insufficient spatio-temporal resolution of the diagnostics prevents further analyses. 
Analyses of the ion temperature avalanche and corrugation~\cite{Ashourvan:2019ek} with the improved diagnostics would be valuable. 
A recent work~\cite{KinSR2023} investigated the hindering role of ion transport events like avalanches in transport barrier formation. 
Secondly, the dissipation mechanism of the $E \times B$ staircase should be studied including a scan of density and toroidal field.
A bifurcation from a dissipative staircase state to a state of staircases interacting broadly in the wavenumber space may be a path for the conventional transport barrier through the staircase merging. 
Thirdly, the high level physics validation using the multi-machine and multi-code data in the near-marginal regime would be desirable. 
%the size scaling study of the $E \times B$ staircase characteristics using the multi-machine data would be desirable. 

\section{Methods \label{sec:methods}}

\subsection{The KSTAR avalanche/staircase plasma \label{sec:avalregime}}

The KSTAR avalanche/staircase plasmas are obtained with the toroidal field $B_\mathrm{T} = 2.7$--$3.0$~T, the plasma current $I_\mathrm{p} = 0.5$--$0.6$~MA and a monotonic $q$ (the central safety factor $q_0 > 1$ and the edge safety factor $q_{95} = 5.6$--$7.0$) with the total $\sim 4$~MW neutral beam injection (NBI). 
Plasmas are attached to the inboard limiter to avoid the transition to the high-confinement mode. 
They can have a tearing mode unstable for the low toroidal field which is then suppressed by the additional 1~MW electron cyclotron resonance heating~\cite{Choi:2019wy}.  

\begin{figure}[t]
\includegraphics[keepaspectratio,width=0.8\textwidth]{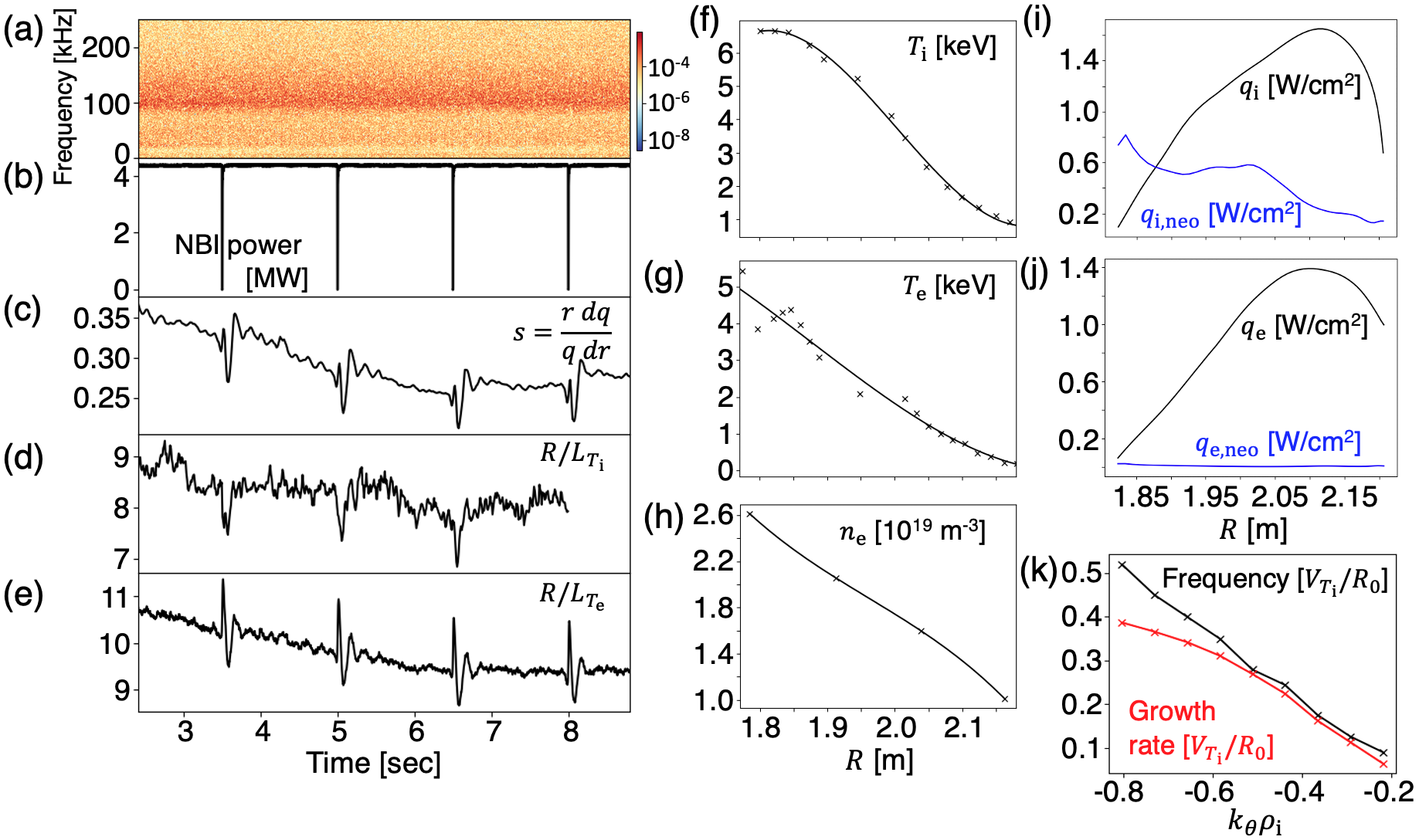}
\caption{(a) Magnetic fluctuation spectrogram. (b) Total NBI power. (c) Magnetic shear $\hat{s}=\frac{r}{q}\frac{dq}{dr}$. (d, e) Normalized inverse temperature gradient scale lengths for ions and electrons $R/L_{T_\mathrm{i,e}}$ with $L_{T_\mathrm{i,e}} = \frac{-T_\mathrm{i,e}}{dT_\mathrm{i,e}/dr}$, respectively. 
(f, g, h) Profiles of ion temperature $T_\mathrm{i}$ and electron temperature $T_\mathrm{e}$ and density $n_\mathrm{e}$ at 4.0~sec, respectively. 
(i, j) Total (black) and neoclassical (blue) ion and electron heat fluxes, respectively. 
(k) The frequency (black) and the growth rate (red) of the linearly most unstable modes near $R_\mathrm{av}$. }
\label{fig:aval2020}
\end{figure}

An example discharge ($B_\mathrm{T} = 2.7$~T, $I_\mathrm{p} = 0.5$~MA and total 4~MW NBI) is introduced in Fig.~\ref{fig:aval2020}. 
Magnetic fluctuation spectrogram in Fig.~\ref{fig:aval2020}(a) shows that MHD instabilities such as sawtooth, tearing modes, or edge localized modes are quiescent.
The total NBI power was kept constant except short blips for diagnostics purposes. 
Time evolutions of typical stability parameters of plasma drift wave instabilities such as magnetic shear $\hat{s}=\frac{r}{q}\frac{dq}{dr}$ and normalized inverse temperature gradient scale lengths $R/L_{T_\mathrm{i,e}}$ with $L_{T_\mathrm{i,e}} = \frac{-T_\mathrm{i,e}}{dT_\mathrm{i,e}/dr}$ are shown, respectively, in Figs.~\ref{fig:aval2020}(c), \ref{fig:aval2020}(d) and \ref{fig:aval2020}(e).
Profiles of ion temperature $T_\mathrm{i}$ and electron temperature $T_\mathrm{e}$ and density $n_\mathrm{e}$ at 4.0~sec are shown, respectively, in Figs.~\ref{fig:aval2020}(f), \ref{fig:aval2020}(g) and \ref{fig:aval2020}(h). 
Total (black) and neoclassical (blue) ion and electron heat fluxes are shown in Figs.~\ref{fig:aval2020}(i) and \ref{fig:aval2020}(j), respectively. 
They are obtained from the power balance analysis using the profiles and TRANSP~\cite{TRANSP} and NUBEAM~\cite{NUBEAM1, GoldstonJCP1981, PankinCPC2004}.  
Note that the ion turbulent heat flux ($q_\mathrm{i} - q_\mathrm{i,neo}$) is comparable to the ion neoclassical heat flux, indicating that the turbulence drive is not too strong in this plasma. 
The linear gyrokinetic simulation was performed using the profiles and the global delta $f$ electrostatic gKPSP code with the bounce-averaged kinetic electron and collisions~\cite{Qi:2016cr, KwonCPC2017}. 
According to the convention in this work, the negative phase velocity of the linearly most unstable modes means that the ion temperature gradient (ITG) mode may be a dominant micro-instability in this plasma.

%The prevalence of the avalanche means that the turbulent transport behavior is non-diffusive, implying that the Kubo number is larger than 1 (see Fig.~\ref{fig:avaleddy}(c)).
%Unfortunately, it is beyond our capability to accurately and properly estimate $\tau_\mathrm{jump}$ and $\tau_\mathrm{int}$ to calculate the experimental Kubo number~\cite{GillotPPCF2023} for the relevant turbulence in the KSTAR core plasma. 

\subsection{Avalanches and the $E \times B$ staircase in KSTAR \label{sec:kstaravst}}

\begin{figure}[t]
\includegraphics[keepaspectratio,width=0.6\textwidth]{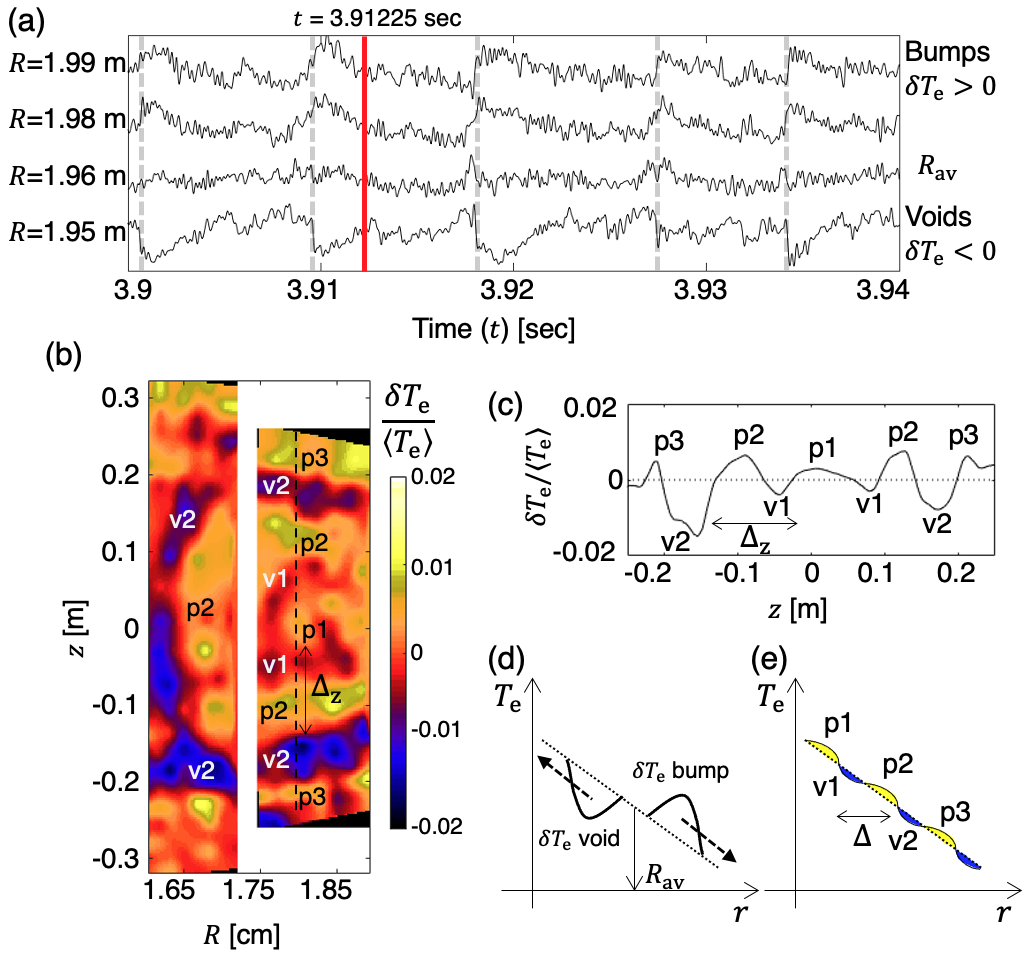}
\caption{(a) $T_\mathrm{e}$ time traces measured at different radial locations (y-axis). Bumps and voids of various size avalanches are generated near $R_\mathrm{av}$ in the plasma core region. Large size avalanches whose bumps propagate to the plasma edge are indicated by dashed lines. (b) Two-dimensional measurement of $\delta T_\mathrm{e} / \langle T_\mathrm{e} \rangle$ at 3.91225~sec in between large avalanches. Jet-like patterns in the $\delta T_\mathrm{e} / \langle T_\mathrm{e} \rangle$ image correspond to a staircase-like $T_\mathrm{e}$ corrugation (the $E \times B$ staircase) and letters ‘p’s and ‘v’s represent peaks and valleys of $\delta T_\mathrm{e} / \langle T_\mathrm{e} \rangle$, respectively. (c) The vertical cut of $\delta T_\mathrm{e} / \langle T_\mathrm{e} \rangle$ measurement along the plasma center. $\Delta_z$ indicates the vertical tread width of the staircase. (d) The illustration of the avalanche, preserving the joint reflection symmetry. (e) The illustration of a staircase-like $T_\mathrm{e}$ profile corrugation with the radial tread width $\Delta$ by the $E \times B$ staircase. Dotted lines in illustrations represent the average temperature $\langle T_\mathrm{e} \rangle$ profile.}
\label{fig:stair2019}
\end{figure}

Transport events of various sizes are observed in a particular regime of KSTAR plasmas without macroscopic MHD instabilities (see Methods~\ref{sec:avalregime}).
Their appearance as bumps and voids on electron temperature ($T_\mathrm{e}$) is detected using the KSTAR electron cyclotron emission (ECE) diagnostics~\cite{KoIEEE2010, YunRSI2014}.
Fig.~\ref{fig:stair2019}(a) shows the filtered $T_\mathrm{e}$ measured at different radial locations in the KSTAR plasmas (major radius $R_0 = 1.8$ m and minor radius $a=0.5$ m~\cite{Park:2019dv}).
Bumps ($\delta T_\mathrm{e} > 0$) propagate down the gradient and voids ($\delta T_\mathrm{e} < 0$) propagate up the gradient against the initiation location $R=R_\mathrm{av}$ ($r/a \sim 0.4$)~\cite{Choi:2019wy}. 
They are called avalanches because they exhibit characteristics of the SOC avalanche in plasmas~\cite{DiamondPoP1995, HahmJKPS2018, SanchezPPCF2015}: (1) the power-law spectrum~\cite{Choi:2019wy}, (2) the large Hurst exponent ($H \sim 0.73 > 0.5$)~\cite{Choi:2019wy}, (3) the joint reflection symmetry~\cite{Choi:2019wy} and (4) the ballistic propagation. 

A staircase-like electron temperature corrugation is observed in the KSTAR avalanche plasmas~\cite{Choi:2019wy}. 
For example, Fig.~\ref{fig:stair2019}(b) shows the two-dimensional measurement of the $T_\mathrm{e}$ corrugation at 3.91225~sec (red line in Fig.~\ref{fig:stair2019}(a)) and its vertical cut is plotted in Fig.~\ref{fig:stair2019}(c). 
The $T_\mathrm{e}$ corrugation appears as poloidally symmetric jet-like patterns in the $\delta T_\mathrm{e} / \langle T_\mathrm{e} \rangle$ image, and its peaks and valleys are indicated by ‘p’s and ‘v’s letters (see Fig.~\ref{fig:stair2019}(e) for the illustration).
This temperature corrugation will be referred to as the $E \times B$ staircase in this paper with caution that the corresponding measurement of the fine scale radial electric field or the poloidal flow variation is left for future work.   
Justification for its naming as the $E \times B$ staircase is given by its behavior and characteristics presented in the main text.
The most notable result is that the corrugation width follows a Fr\'echet distribution as the $E \times B$ staircase width in the simulation~\cite{DifNF2017, MilovanovPRE2021}. 

\subsection{Diagnostics and data analysis schemes \label{sec:diagmeth}}

Ion and electron temperature profiles are obtained from the charge exchange recombination spectroscopy~\cite{KoIEEE2010} and one-dimensional calibrated electron cyclotron emission (ECE) diagnostics~\cite{Jeong:2010bq} in the low field side $R>R_0$, respectively.
The density profile is reconstructed from the interferometer with multiple lines of the sight~\cite{Lee:2016fw}. 

The one-dimensional ECE diagnostics, the two-dimensional electron temperature fluctuation diagnostics (electron cyclotron emission imaging, ECEI~\cite{YunRSI2014}), and the two-dimensional density fluctuation diagnostics (beam emission spectroscopy, BES~\cite{Nam:2014kd}) are utilized to detect and investigate avalanches.
The large avalanche can be identified by finding peaks or valleys in the low-pass filtered $T_\mathrm{e}$ time traces over the radial region.
The maximum correlation time lag between $T_\mathrm{e}$ data from spatially separated channels is used to measure the propagation speed of the large avalanche.  
The cross power and cross phase spectra~\cite{Choi:2019tw} between data from two adjacent channels are used to obtain the spectra of turbulence eddies ($|\delta n_\mathrm{e} / \langle n_\mathrm{e} \rangle|^2$) or the avalanche pseudo-size ($|\delta T_\mathrm{e} / \langle T_\mathrm{e} \rangle|^2$) and to estimate the phase velocity of turbulence eddies, respectively. 
 
\begin{figure}
\includegraphics[keepaspectratio,width=0.8\textwidth]{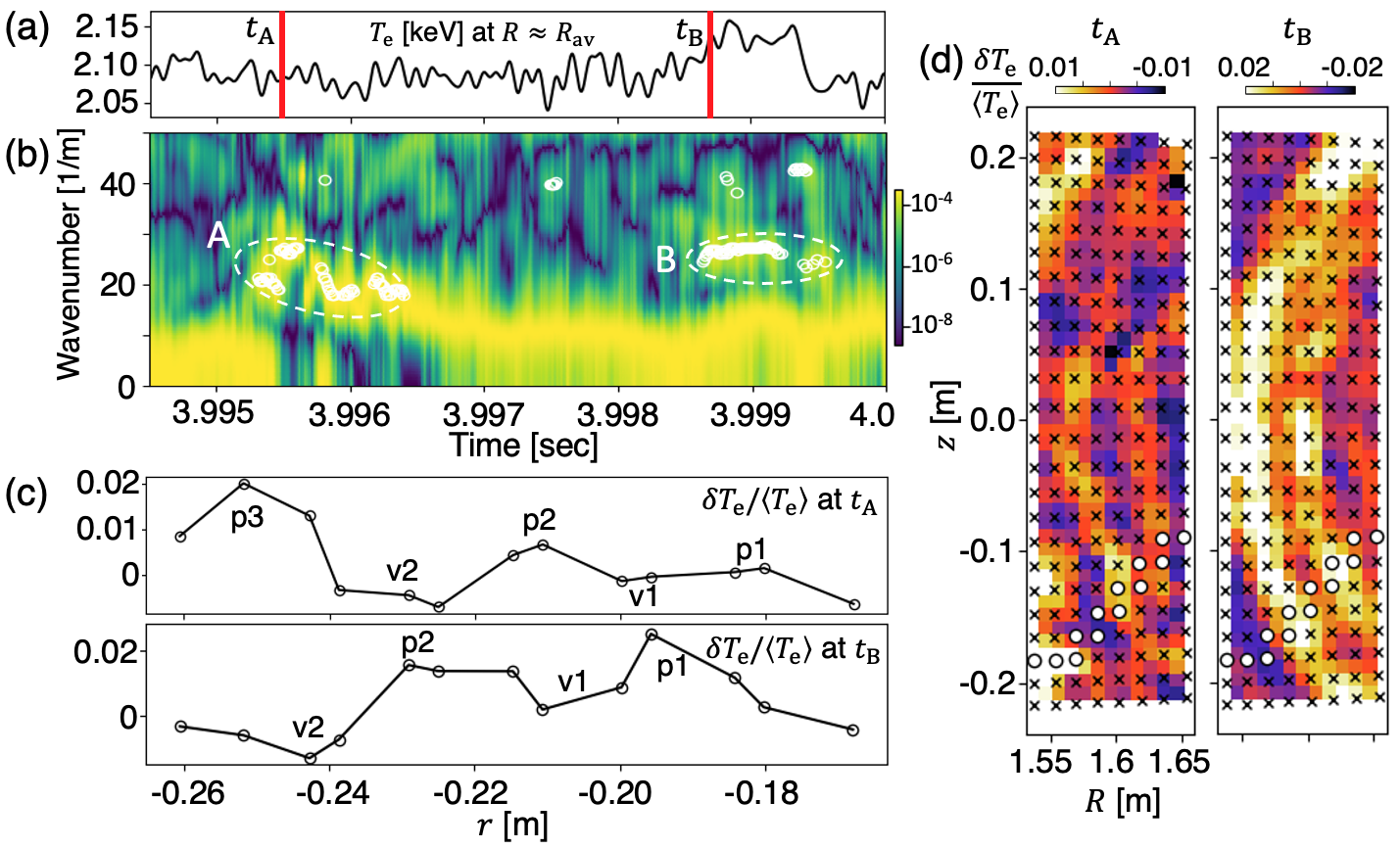}
\caption{(a) $T_\mathrm{e}$ time trace at $R \approx R_\mathrm{av}$. (b) The radial wavenumber spectrogram obtained using the successive Lomb-Scargle periodograms of the non-uniform radial $\delta T_\mathrm{e} / \langle T_\mathrm{e} \rangle$ profiles. (c, d) The radial $\delta T_\mathrm{e} / \langle T_\mathrm{e} \rangle$ profiles and the $\delta T_\mathrm{e} / \langle T_\mathrm{e} \rangle$ images at times $\tau_\mathrm{A}$ and $\tau_\mathrm{B}$, respectively. }
\label{fig:stair2020}
\end{figure}

For the detection of the $E \times B$ staircase and the measurement of its characteristics, the high resolution $\delta T_\mathrm{e} / \langle T_\mathrm{e} \rangle$ data from the ECEI diagnostics~\cite{YunRSI2014} in the high field side $R<R_0$ are utilized.
The detailed procedure is explained as follows with two representative cases. 
Fig.~\ref{fig:stair2020}(a) shows the $T_\mathrm{e}$ trace measured near $R \approx R_\mathrm{av}$ for some period. 
A bump of the large avalanche was observed around 3.999~sec.
Some of the ECEI channels with good signal-to-noise ratio, indicated by white-filled circles in Fig.~\ref{fig:stair2020}(d), are selected to reconstruct a radial $\delta T_\mathrm{e} / \langle T_\mathrm{e} \rangle$ profile along the minor radius ($r$) on the midplane ($z=0$) as shown in Fig.~\ref{fig:stair2020}(c). 
Their measurements on ($R, z$) space are projected on the midplane following the magnetic flux surface shape of EFIT calculation~\cite{Lao:1985hn}, which provides an enhanced effective radial resolution~\cite{Choi:2014kj}. 
The channel position of the ECEI diagnostics was calculated using the forward modeling of the ECE transport~\cite{syndia}.
The raw ECEI data was filtered by the boxcar averaging with the averaging window size of 300~$\mu$s (this boxcar averaging window size is taken as the minimum lifetime for the meaningful detection). 
Then, the Lomb-Scargle periodogram~\cite{VanderPlas2018} is used to estimate the radial wavenumber ($k=1/\Delta$) spectrum for the radially non-uniform $\delta T_\mathrm{e} / \langle T_\mathrm{e} \rangle$ measurement. 
The resulting radial wavenumber spectrogram with the successive Lomb-Scargle periodograms is shown in Fig.~\ref{fig:stair2020}(b). 
Empty white circles in Fig.~\ref{fig:stair2020}(b) indicate significant peaks of the radial wavenumber periodogram whose power and also wavenumber are higher than some thresholds. 
Low radial wavenumber ($k < 18$~m$^{-1}$) peaks are ignored because they are indistinguishable from the trails of the large avalanche in the limited diagnostics view in the experiment. 
Next, a cluster identification algorithm~\cite{OPTICS, SKLEARN} is used to group close measurements (the temporal separation $< 300$~$\mu$s~$=$ the boxcar averaging window size; the radial width ($\Delta$) separation $< 2$~cm). 
Clusters with sufficient lifetime ($\ge$~300~$\mu$s) are counted only to avoid possible temporary noise contribution. 
In Fig.~\ref{fig:stair2020}(b), the $E \times B$ staircase A and B are found, and A can be classified as the one away from the large avalanche (class A) and B as the other near the large avalanche within 2~msec (class B). 
%This classification is found to be important since they have different $\Delta$ distributions, possibly resulting from the avalanche contribution. 
Radial $\delta T_\mathrm{e} / \langle T_\mathrm{e} \rangle$ profiles of A and B at times $t_\mathrm{A}$ and $t_\mathrm{B}$ are shown in Fig.~\ref{fig:stair2020}(c), and two-dimensional $\delta T_\mathrm{e} / \langle T_\mathrm{e} \rangle$ measurements at those times are shown in Fig.~\ref{fig:stair2020}(d). 
Values from bad signal-to-noise ratio channels were interpolated in Fig.~\ref{fig:stair2020}(d). 
Peaks and valleys of temperature corrugations are indicated by letters ‘p’s and ‘v’s, respectively, and jet-like patterns appear. 
The staircase B co-exists with the growing bump of the avalanche which corresponds to the low wavenumber component. 
The lifetime ($t_\mathrm{life}$), the power ($|\delta T_\mathrm{e} / \langle T_\mathrm{e} \rangle|^2$), and the radial tread width ($\Delta$) of the staircase are taken by the temporal width, the average power, and the inverse of the average wavenumber ($\Delta = 1/k$) of the clustered measurements, respectively.

\section*{Acknowledgments}

One of the authors (M.J.C.) acknowledges helpful discussions with K. Razumova, Y. Kishimoto, Y. Kosuga, W. Wang, K. Ida, T. Kobayashi, F. Kin and M. Sasaki during the UNIST-Kyoto University workshop, the Asia-Pacific Transport Working Group meeting, the Asia-Pacific Conference on Plasma Physics and elsewhere.
This work was supported by R\&D Programs of ``KSTAR Experimental Collaboration and Fusion Plasma Research (EN2401-15)'' and ``High Performance Fusion Simulation R\&D (EN2341-9)'' through Korea Institute of Fusion Energy (KFE) funded by the Government funds and by the NRF of Korea under grant No. RS-2023-00281272. 
Computing resources were provided on the KFE computer, KAIROS, funded by the Ministry of Science and ICT of the Republic of Korea (EN2441-10).
The author (M.J.C.) would like to thank the Isaac Newton Institute for Mathematical Sciences, Cambridge, for support and hospitality during the programme Layering — A structure formation mechanism in oceans, atmospheres, active fluids and plasmas, where work on this paper was partially undertaken. 
This work was supported by EPSRC grant EP/R014604/1.

\section*{Competing interests}

The authors declare no competing interests.

\section*{Data availability}

Data are available from the corresponding author upon request.

%\section*{Code availability}
%
%The codes used for figures of this article are available via GitHub repositories~\cite{Choi:2019tw, fluctana, syndia}. 

\section*{References}

\bibliographystyle{naturemag}
%\bibliography{staircase}

\end{document}